\documentclass[aps,prb,superscriptaddress,
 amsmath,amssymb,
 reprint,%
]{revtex4-2}

\usepackage{slantsc}
\usepackage{lmodern}
\usepackage{amsmath}
\usepackage{amssymb}   
\usepackage{xcolor}
\usepackage{graphicx}
\usepackage{dcolumn}
\usepackage{bm}
\usepackage{float}

\usepackage[utf8]{inputenc}
\usepackage[T1]{fontenc}
\usepackage{mathptmx}

\begin{document}


\title{Atomistic tight-binding Hartree-Fock calculations of multielectron configurations\\ in P-doped silicon devices: wavefunction reshaping}

\author{Maicol A. Ochoa}
\email{maicol@nist.gov}
\affiliation{National Institute of Standards and Technology, Gaithersburg, MD 20899, USA}
\affiliation{Department of Chemistry and Biochemistry, University of Maryland, College Park, MD, 20742, USA}

\author{Keyi Liu}
\affiliation{National Institute of Standards and Technology, Gaithersburg, MD 20899, USA}
\affiliation{Joint Quantum Institute, University of Maryland, College Park, MD, 20742, USA}

\author{Piotr R\'o\.za\'nski}
\affiliation{Institute of Physics, Faculty of Physics, Astronomy and Informatics, Nicolaus Copernicus University, ul. Grudziadzka 5,87-100 Toru\'n, Poland}

\author{Michal Zieli\'nski}
\affiliation{Institute of Physics, Faculty of Physics, Astronomy and Informatics, Nicolaus Copernicus University, ul. Grudziadzka 5,87-100 Toru\'n, Poland}

\author{Garnett W. Bryant}
\email{garnett.bryant@nist.gov}
\affiliation{National Institute of Standards and Technology, Gaithersburg, MD 20899, USA}
\affiliation{Joint Quantum Institute, University of Maryland, College Park, MD, 20742, USA}

\date{\today}

\begin{abstract}

  Donor-based quantum devices in silicon are attractive platforms for universal quantum computing and analog quantum simulations. The nearly-atomic precision in dopant placement promises great control over the quantum properties of these devices. We present atomistic calculations and a detailed analysis of many-electron states in a single phosphorus atom and selected phosphorus dimers in silicon. Our self-consistent method involves atomistic calculations of the electron energies utilizing representative tight-binding Hamiltonians, computations of Coulomb and exchange integrals without any reference to an atomic orbital set, and solutions to the associated Hartree-Fock equations. First, we assess the quality of our tight-binding Hartree-Fock protocol against Configuration-Interaction calculations for two electrons in a single phosphorus atom, finding that our formalism provides an accurate estimation of the electron-electron repulsion energy requiring smaller computational boxes and single-electron wavefunctions. Then, we compute charging and binding energies in phosphorus dimers observing their variation as a function of impurity-impurity separation. Our calculations predict an antiferromagnetic ground state for the two-electron system and a weakly bound three-electron state in the range of separations considered. We rationalize these results in terms of the single-electron energies, charging energies, and the wavefunction reshaping.
\end{abstract}

\maketitle

\section{Introduction}

Silicon nanostructures that incorporate dopants forming arrays with atomic precision are promising platforms for developing quantum simulators, quantum materials, and other quantum technologies. Soon after the proposal by Kane\cite{kane1998silicon} of a silicon-based quantum computer encoding information in the nuclear spin of $^{31}$P donor atoms, and the implementation proposed by Loss and DiVincenzo\cite{loss1998quantum} based on the spin states of coupled single-electron quantum dots, initial attempts to fabricate impurity-based devices with the required atomic precision appeared in the literature\cite{o2001towards, pla2013high, dehollain2014single,schmucker2019low, wyrick2022enhanced}. This, eventually, led to the fabrication of the single-atom transistor\cite{fuechsle2012single}, and other devices comprising a few phosphorus atoms\cite{buch2013spin,wyrick2019atom, stock2020atomic,ivie2021impact,campbell2023quantifying}. Recently, reports on the analog quantum simulations of the many-body Su-Schrieffer-Heeger (SSH) model\cite{kiczynski2022engineering}, and the $3 \times 3$ extended Fermi-Hubbard model\cite{wang2022experimental} in dopant-based silicon devices appeared.

Electronic structure calculations of bound electrons in impurity-based devices in semiconductors require efficient methods that can account for the device's atomistic details, including the impurity array confinement, the host material electronic properties, potential energy variations along the device resulting from external gates and, importantly, the electron-electron interactions. This characterization is critical to understand the performance, operation and potential of these devices in applications such as quantum simulations of the Fermi-Hubbard model. A typical dopant array may cover a few nanometers, resulting in electronic structure calculations that must incorporate the multi-million host-material atoms surrounding the array. Large computational boxes extending beyond the array's domain, are also needed to account for the extended structure of the host material and avoid artificial distortions due to the computational boundary. At present, such types of calculations are forbidden for the most accurate electronic structure methods, and are limited a few thousand atoms\cite{jia2024atomistic}. Thus, it is common practice to use semi-empirical tight-bind methods (TB) to describe these systems. For the case of phosphorus-atom arrays in silicon, tight-binding calculations provide a complete picture of the electron density of a single-electron in dopant arrays\cite{ochoa2024single,liu2023} \textcolor{black}{ at the same time that it reproduces the relevant bandgaps and effective masses of the host material}. One can solve the coupled Poisson-Schr\"odinger equations\cite{trellakis2005efficient,ryu2015tight} or implement configuration-interaction (CI) calculations\cite{munia2024superexchange} starting from the TB single-particle basis to obtain the energy and structure of multielectron states in the dopant array. As is the case with most common algorithms for many-body quantum methods, the CI computational complexity grows exponentially with electron number and single-particle basis size, limiting the efficient implementation of this approach to few-electron systems. On the other hand, Poisson-Schr\"odinger solvers require semiempirical potentials to account for electron exchange and correlation, which should be optimized for the specific system; and may incorporate unphysical electron self-interactions.

In this paper, we present a formalism for calculating many-body electron properties in dopant arrays that combines atomistic tight-binding calculations and the Hartree-Fock approximation. This method solves the associated interaction integrals without utilizing explicit orbital wavefunctions for the local atomic orbitals, needing only the local tight-binding amplitudes. Specifically, our method self-consistently finds the single-electron states of the tight-binding Hamiltonian that include electron-electron repulsion energy potentials and single-particle states that solve the associated Hartree-Fock equations. \textcolor{black}{Each iteration generates a new set of tight-binding amplitudes that account for mutual electron-electron interaction, removing self-interactions while incorporating exchange in a new set of amplitudes representin the Hartree-Fock states. Tight-binding parameters, such as orbital and tunneling energies do not change between iterations.} In this form, the self-consistent approach results in electron wavefunctions reshaped by electron-electron interactions, which form the many-body state for the multi-electron system as the Slater determinant of these single-particle Hartree-Fock states. We demonstrate the advantage of this approach in obtaining the many-body energy and the reshaped electron distribution over direct CI calculations, by computing the charging and binding energies for a second electron in a single phosphorus dopant in silicon. \textcolor{black}{Our calculations treat explicitly the additional electrons bound to the impurity, relative to the electronic structure of a single silicon atom,  while treating the remaining electrons within the tight-binding frame.} Next, we apply our formalism to systems consisting of two phosphorus atoms at short separations, forming atomic clusters. Multiple dopants can form highly confined regions in the neighborhood of closely packed dopants. Conversely, when the impurity-impurity separation is large, dopants form arrays of tunnel-coupled quantum wells. After analyzing the energy distribution of the valley-split lowest single-electron states in these clusters, we report binding and charging energies for one, two, and three electrons. Our result suggests that the impurity-impurity separation in a dimer cluster is a good measure of the electron confinement in clusters, revealing simple trends in the binding and charging energy for two and three electrons. Next, we consider phosphorus dimers at larger separations along the [100] direction, transitioning from a cluster to an array configuration. We note that two-electron binding and charging energies decay with separation, favoring an antiferromagnetic ground state in all cases. The dimer can hold an additional electron, forming a structure with a charge of -1. This state is weakly bound, and their binding energy is very sensitive to the impurity-impurity separation.

The organization of the paper is as follows. In Sec.\ \ref{sec:theory} we describe in detail the self-consistent protocol for calculating the many-body electron properties by combining tight-binding calculations and the Hartree-Fock approximation. Then, in Sec.\ \ref{sec:1P} we study the two-electron problem in a single phosphorus atom in Silicon, calculating charging and binding energies. In Sec. \ref{sec:2Pclusters} we investigate multielectron states in two-phosphorus-atom clusters and analyze their variation with impurity separation. Next, in Sec.\ \ref{sec:2Parray}, we consider one, two, and three electron states in phosphorus dimers at separations larger than 1 nm.  Finally, we summarize in Sec.\ \ref{sec:conclusion}.

\section{TB-HF formalism}\label{sec:theory}

In this section, we introduce the Tight-Binding Hartree-Fock formalism (TB-HF), as an efficient method to calculate many-electron states and energies in impurity arrays/clusters in silicon. The protocol combines exact electronic tight-binding calculations\cite{boykin2004valence,ochoa2024single} and solutions to the associated Poisson equations to obtain the electric potential and charge distribution in the impurity array in a self-consistent form. The TB-HF method for an $N$-electron system consists of three cyclic steps:

\begin{enumerate}
\item {\sl \underline{TB calculation} } We carry out tight-binding calculations to obtain single-electron states (TB-states) and energies (TB-energies) of the impurity array. We include the screening potential energy $J_N$ due to the electron-electron repulsion potential obtained at step 3.
\item {\sl \underline{HF equations} } We write the Hartree-Fock equations in the TB-state basis in step 1 and solve these equations obtaining Hartree-Fock single-electron state (HF-states) and energies (HF-energies).  The many-electron state is the Slater determinant of the $N$ lowest HF-states. 
\item {\sl\underline{Total electron-electron potential}} From the many-body wavefunction obtained in step 2, we obtain the total $N$-electron density and the corresponding total electrostatic potential $J^{(N)}$ by solving the associated Poisson equation.
\end{enumerate}

The protocol runs in a loop until one achieves self-consistency in the potential $J^{(N)}$ or the electron density. It can be implemented to calculate the $N$-electron ground-state energy, as well as excited energy states with a restricted spin configuration, such as triplet states. We now describe each step in more detail.
 
We implement the empirical $sp^3d^5s^*$ atomistic tight-binding model with spin for Si utilizing the parameter set in Ref.\ \citenum{boykin2004valence}, and represent each phosphorus atom in the array/cluster by a screened Coulomb potential centered at the impurity position\cite{kohn1955theory,luttinger1955motion,ochoa2024single}, with a central cell correction. Specifically, each phosphorus atom in the array
replaces a Si atom, introducing a confinement potential for
the additional electron that we model as a screened Coulomb
potential,
\begin{equation}
U_P(r) =
\begin{cases}
\frac{-e}{ 4 \pi \varepsilon_{\rm Si} |\vec{r} -\vec{r}_P|}&  \vec{r} \neq \vec{r} _P\\
U_{\rm CCC}& \vec{r} = \vec{r}_P
\end{cases}
\end{equation}
where $\varepsilon_{\rm Si}$ is the silicon dielectric constant, $\vec{r}_P$ is the impurity location, and $U_{\rm CCC}$ is the central cell correction.

The TB calculation includes the potential energy $J^{(N)}(\vec{r}) = \sum_{i=1}^N J_i(\vec{r})$ corresponding to the total electron-electron repulsion for $N$ electrons in the array/cluster. We can initialize our calculation by assuming that this potential is zero or estimate it as the sum of the electron density for single electron states in each individual P atom. Thus, we write the tight-binding Hamiltonian $H_{\rm TB}$
\begin{align}
  \hat H_{\rm TB}(\vec{r}) = & \hat H_o(\vec{r}) + \hat J^{(N)}(\vec{r}) \label{eq:HTB}\\
  \hat H_o (\vec{r}) = & \hat H_{\rm Si}(\vec{r})+ \hat U_{\rm P-array}(\vec{r}), \label{eq:Ho}       
\end{align}
where $\hat H_{\rm Si}$ is the Hamiltonian for the silicon matrix\cite{boykin2004valence}, and $\hat U_{\rm P-array}(\vec{r}) = \sum_i^N \hat U_{P_i}(\vec{r})$ is the screened nuclear potential of the impurity array.  We solve the eigenvalue \textcolor{black}{matrix} equation, $\hat H_{\rm TB} | \phi_k \rangle = \varepsilon_k^{\rm TB} | \phi_k \rangle$, and find states  $| \phi_k \rangle$ with energies $\varepsilon_k^{\rm TB}$ falling in the band-gap and near the conduction band edge. 

 Starting from the $N$-electron Hamiltonian
\begin{align}
  \hat H_{(N)} (\vec{r}_1, \dots, \vec{r}_N) =& \sum_{k=1}^N \hat H_o (\vec{r}_k) +\sum_{k=1}^{N-1} \sum_{l>k}^N \frac{1}{|\vec{r}_k- \vec{r}_l|}, \label{eq:HN}
\end{align}
we write the associated self-consistent HF equations\cite{szabo2012modern}
\begin{align}
  \varepsilon^{\rm HF}_i | \psi_i (\vec{r}_1) \rangle =&  \hat H_o(\vec{r}_1)| \psi_i(\vec{r}_1) \rangle \notag \\
& \hspace{1cm}  + \sum_{j\neq i } \hat J_j(\vec{r}_1) | \psi_i (\vec{r}_1) \rangle - \hat K_{ij}(\vec{r}_1) | \psi_j (\vec{r}_1) \rangle  \\
  =& \hat H_{\rm TB}(\vec{r}_1)| \psi_i(\vec{r}_1) \rangle \notag \\
  &\hspace{1cm}-  \hat J_i(\vec{r}_1) | \psi_i (\vec{r}_1) \rangle - \sum_{j\neq i } \hat K_{ij}(\vec{r}_1) | \psi_j (\vec{r}_1) \rangle \label{eq:HNtb}
\end{align}
in terms of the Hartree-Fock single-electron wavefunction $|\psi_i \rangle $, with the Coulomb and exchange potentials given by
\begin{align}
  \hat J_j(\vec{r}_1) =& \langle \psi_j (\vec{r}_2) | \frac{1}{|\vec{r}_2- \vec{r}_1|} | \psi_j (\vec{r}_2) \rangle \label{eq:Jj}\\
  \hat K_{ij}(\vec{r}_1) =&\langle \psi_j (\vec{r}_2) | \frac{1}{|\vec{r}_2- \vec{r}_1|} | \psi_i (\vec{r}_2) \rangle. \label{eq:Kij}
\end{align}
We then assume the set of TB states $\{ | \phi_k \rangle \}$ forms a complete basis in which we can expand the HF states\cite{szabo2012modern} $| \psi_i \rangle = \sum_{k=1}^M c_i^{(k)} | \phi_k \rangle$, for a sufficiently large $M \geq N$.  We write the canonical Hartree-Fock equations on this basis and solve the matrix resulting matrix equation $ \mathbf{H C} = \mathbf{S C E} $, with
\begin{align}
    \mathbf{H} =& \hat H_o - \hat F \label{eq:Hhf}\\
    [\hat H_o]_{ij} =& \delta_{ij} \varepsilon_i^{\rm TB}\\
    [\hat F]_{ij} =& \langle \phi_i | \hat K_{ji} | \phi_j \rangle \label{eq:Fij},
\end{align}
obtaining the HF-energies $\{\varepsilon_i^{HF}\}$ and coefficients $\{c_i^{k}\}$  defining the HF-states in terms of $M$ TB states. We remark that the matrix $\hat F$ in Eq.\ \eqref{eq:Fij}, has diagonal elements of the form $\langle \phi_j | \hat J_j | \phi_j \rangle$, corresponding to the self-interactions introduced during the TB-calculation and that are removed in Eq.\ \eqref{eq:Hhf}.   The many-body wavefunction is the antisymmetric $N$-electron Slater determinant
\begin{equation}
  \label{eq:SlaterPsi}
  |\Psi^{N} \rangle = \frac{1}{\sqrt{N!}} \sum_{s \in \mathit{S}_N}(-1)^{{\rm sgn}(s)}|\psi_{s(1)}(\vec{r}_1) \dots \psi_{s(N)}(\vec{r}_N)\rangle.
\end{equation}
During the calculation, and by virtue of the identity
\begin{align}
  \nabla_1^2  \frac{1}{|\vec{r}_2- \vec{r}_1|} = - 4 \pi \delta(\vec{r}_1 -\vec{r}_2),
\end{align}
we determine the potentials in Eqs.\ \eqref{eq:Jj} and \eqref{eq:Kij} by solving the associated differential equations
\begin{align}
  \nabla_1^2 \hat J_j (\vec{r}_1) =& - \left| \psi_j(\vec{r}_1) \right |^2  \label{eq:PoisJ}\\
  \nabla_1^2 \hat K_{ij} (\vec{r}_1) =& - \psi_j(\vec{r}_1)^* \psi_i (\vec{r}_1) ,\label{eq:PoisK}
\end{align} 
with boundary conditions obtained from the asymptotic behavior of Eqs.\ \eqref{eq:Jj} and \eqref{eq:Kij} as $\vec{r}_1$ approaches to the computational boundary box $\Omega$. Having obtained the many-body wavefunction $| \Psi ^N \rangle$ and the potentials $\{ J_i\}$, we recalculate $J^{(N)}$ and repeat the process until we achieve self-consistency in the electron potential or density. This completes the protocol.

Finally, we make the following remarks about the TB-HF method introduced in this section. First, the present protocol does not require specifying the $sp^3d^5s^*$ atomic orbitals, as it relies only on the tight-binding amplitudes obtained in the first step to calculate potentials in Eqs.\ \eqref{eq:PoisJ} and \eqref{eq:PoisK}. Indeed, we obtain the densities $ \left| \psi_j(\vec{r}_1) \right |^2$ and products $ \psi_j(\vec{r}_1)^* \psi_i (\vec{r}_1)$ from the tight-binding amplitudes and numerically solve for $\hat J_i$ and $\hat K_{ij}$. We also obtain in this form the matrix elements entering in the generalized eigenvalue problem defined by the Hartree-Fock approximation. Second, we remind the reader that the HF approach naturally introduces exchange into the calculation of the many-body states, removing also unphysical self-interactions. This eliminates the need to introduce additional potentials into the self-consistent calculation to account for exchange, as it is frequently done in self-consistent solutions of the tight-binding model and associated Poisson equations \cite{lee2011electronic,weber2012ohm,ryu2013atomistic,ryu2015tight,donnelly2023multi}. We will also find below that by introducing electron-electron repulsion potentials in step 1, we effectively reshape the single-electron wavefunctions and, therefore, the electron density. As compared with CI theory\cite{munia2024superexchange}, the TB-HF approach does not include correlation energy, but one can run an additional CI calculation, starting from the HF-state basis, to account for this correction. We denominate such an approach as the TB-HF-CI$_o$ method.  \textcolor{black}{The latter includes correlation on a basis that satisfies the Hartree-Fock equations.}

\section{ The Phosphorus anion P$^-$}\label{sec:1P}

In this section, we study the electronic structure of a single phosphorus atom in silicon with charge -1. Within the semiempirical tight-binding model for the silicon matrix\cite{ochoa2024single}, this problem corresponds to solving the two-electron problem in the screened spherical potential defined by the P atom. For this system, experimental bulk measurements\cite{taniguchi1976d,norton1976photoconductivity} report a charging energy of 43.6 meV as well as a binding energy of about 2 meV for the second electron, indicating that this is a stable, weakly bound state. Moreover, recent studies in the single-atom transistor\cite{fuechsle2012single} reveal the existence of a stable  P$^{-1}$ state in the charge stability diagrams.   Theoretical studies based on configuration-interaction calculations\cite{ryu2015tight,tankasala2018two,rozanski2023exploiting} reproduce these charging energies with a significant computational cost.
 Here, we implement the TB-HF method introduced in Sec.\ \ref{sec:theory}, utilizing Coulombo\cite{rozanski2023exploiting} -- an efficient, parallelized routine with quasilinear scaling of computational time that calculates Coulomb matrix elements within the tight-binding framework--  to obtain Coulomb and exchange integrals, and electron-electron repulsion potentials. We demonstrate that we can achieve good agreement with the expected values with a lower computational cost.

\begin{figure}[t]
  \centering
  \includegraphics[scale=0.37]{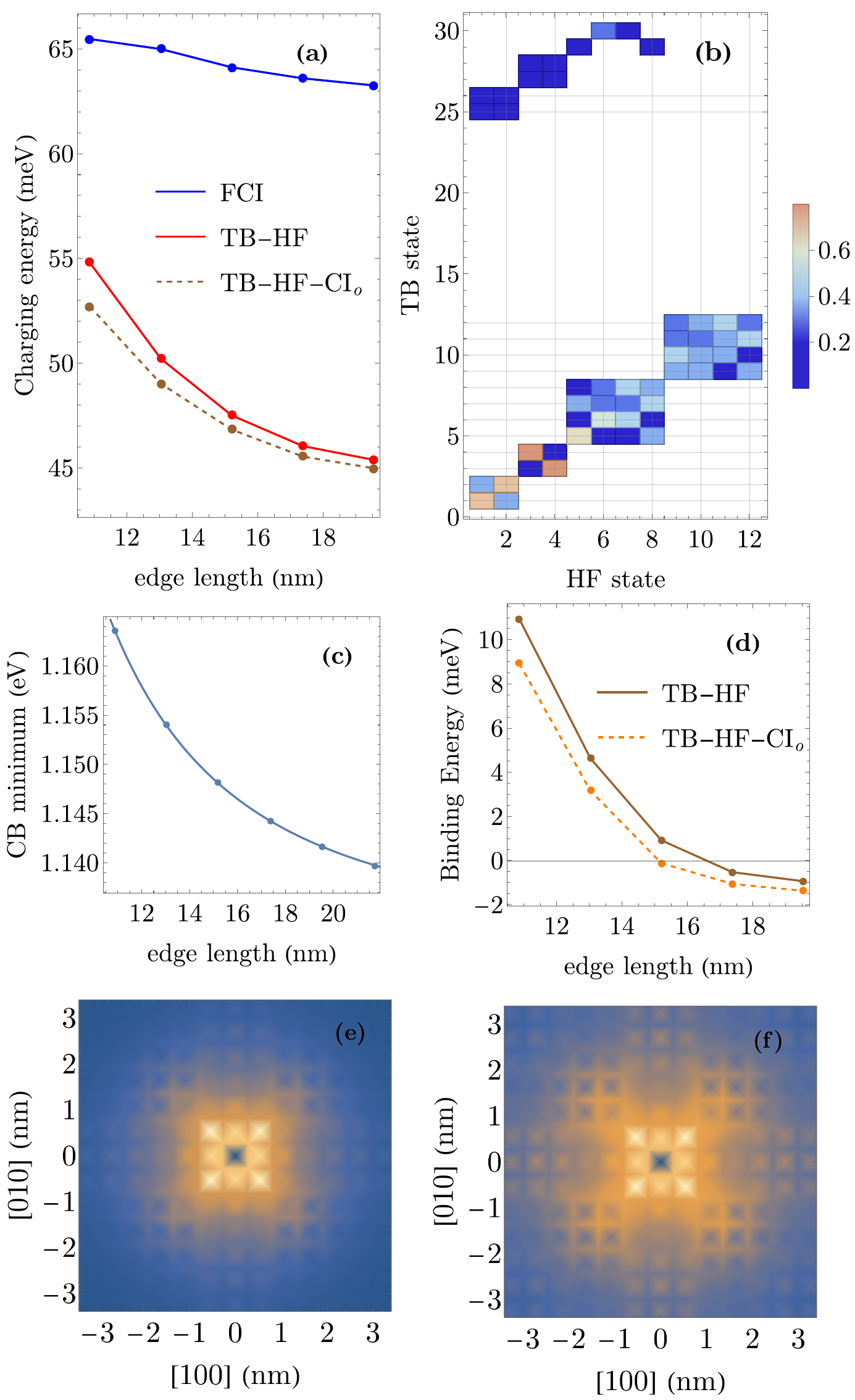}
  \caption{The phosphorus anion P$^{-1}$ in silicon. a) Charging energy for the second electron obtained from a Configuration Interaction (blue), Hartree-Fock (red), and combined Hartree-Fock Configuration Interaction (brown, dashed) calculations as a function of the computational box edge length b) Overlap $|\langle \phi_i^{\rm TB} | \phi_j^{\rm HF} \rangle |^2 $ between the single-electron states used in the CI calculation $| \phi_i^{\rm TB} \rangle$ and the single electron basis obtained self-consistently during the HF calculation  $| \phi_j ^{\rm HF}$, with $1 \le i \le 36$ and $1 \le j \le 12$. The overlaps show that the HF calculation captures self-consistently the reshaping of the wavefunction (c) Conduction band edge as a function of the computational box edge length. The fit line displayed is . (d) Binding energy for corresponding to the transition $P^{-1} \to P^0 + e({\rm CB}(\infty))$. (e) Radial distribution function \textcolor{black}{slice} for the single-electron ground state $| \phi_1^{\rm TB} \rangle$ \textcolor{black}{over the [001] direction}. (f) Radial distribution function slice for the single-electron Hartree Fock state $| \phi_1^{\rm HF} \rangle$ \textcolor{black}{over the [001] direction}. } 
  \label{fig:1p2eHF}
\end{figure}

As a result of the electron interactions, the $P^{-1}$ electronic energy is larger than the sum of the single-electron energies of each electron in the non-interacting ground state. We define the charging energy (CE) for this system  by
\begin{equation}
  {\rm CE}= E_{P^{-1}} - 2 E_{P^0},\label{eq:CE1P}
\end{equation}
where $E_{P^{n}}$ is the electronic energy for negatively charged ($n=-1$) and neutral ($n=0$) phosphorus atom in Si. \textcolor{black}{We remark that we only treat the explicitly the additional electrons bound to the impurity relative to a silicon atom, that is, two electrons in the case of $P^{-1}$ and one in the case of $P^{0}$}. In this form, the charging energy is a measure of interaction energy due to the electron-electron repulsion and wavefunction reshaping, extra energy needed by a second electron to form the $P^{-1}$ state.  In Fig.\ \ref{fig:1p2eHF}a, we report the CE as a function of computational box size, calculated by three different methods: the full configuration interaction (FCI), the TB-HF, and the TB-HF-CI$_o$ introduced here. We used the lowest 36 TB states in the three calculation types as a basis, which include the twelve valley-split spin-degenerate 1$s$ states characteristic of a single electron in a P atom in Si. The others are higher states and can include conduction band states. Since the second electron is only weakly bound to the P atom, the calculation of the multi-electron energy is sensitive to the conduction band edge and can vary with the computational box size\cite{tankasala2018two,rozanski2023exploiting}. As evidenced by Fig.\ \ref{fig:1p2eHF}a, the HF-TB method approaches faster the expected value \cite{taniguchi1976d,norton1976photoconductivity} of 43.6 meV for the CE than the full configuration interaction calculation, providing a better estimate of the charging energy in smaller computational boxes. Indeed, by including the conduction band and higher energy states in the basis, the FCI calculation broadens the electron distribution, compensating for the mutual repulsion. Accurate higher bound states, such as $2s$ states, as well as conduction band states, demand large computational boxes as these are only weakly confined. On the contrary, the self-consistent TB-HF protocol reshapes the wavefunction on each iteration, converging to a broadened self-consistent HF-basis that more efficiently captures contributions from higher energy states to the many-body wavefunction. Moreover, we note that including electron correlation energy results in a minor but additional improvement of the binding and charging energies. Overall, we can conclude that wavefunction reshaping is the physical mechanism that allows for a weakly bound two-electron state. To illustrate this point, we first show in Fig.\ \ref{fig:1p2eHF}b the overlap between the single-particle basis used in the FCI calculation and the self-consistent HF states. We note that the lower two HF states -- which form the many-body wavefunction for the two-electron system as in Eq.\ \eqref{eq:SlaterPsi} -- overlap with the higher energy states that contribute significantly to the FCI wavefunction. We ascribe these overlapping TB-states, listed as 25 and 26 in Fig.\ \ref{fig:1p2eHF}b, to the 2$s$ manifold for a single electron in a P atom. Second, we calculate in Figs.\ \ref{fig:1p2eHF}(e) and (f) the corresponding radial distribution function 
\begin{equation}
    \label{eq:rad}
    g(\vec{r},\vec{r}_o) = |\vec{r}-\vec{r}_o| |\phi(\vec{r})|^2,
\end{equation}
 for the single-electron ground state in a phosphorus atom $| \phi_1^{\rm TB} \rangle$, and the lowest HF state $| \phi_1^{\rm HF} \rangle$. We show the radial distribution function in the (100) plane containing the P atom, setting $\vec{r}_o$ as the impurity position $\vec{r}_P$. Thus, since the electron density decays fast far from P, $g(\vec{r},\vec{r}_P)$ vanishes when $\vec{r} = \vec{r}_P$ and when $|\vec{r}-\vec{r}_P|$ is large. By comparing Figs.\  \ref{fig:1p2eHF}(e) and (f), we find that the electron density in the HF state $| \phi_1^{\rm HF} \rangle$ is wider than the single electron density $| \phi_1^{\rm TB} \rangle$, demonstrating the rescaling achieved by the TB-HF method. In fact, the mean radial dispersion 
 \begin{equation}
 \label{eq:mrd}
     \langle r \rangle = \int_{\rm BOX} g(\vec{r},\vec{r}_P) \, dx\, dy \, dz ,
 \end{equation}
 is much larger for the HF -state ($\langle r \rangle_{\rm HF} = 3.58 $ nm) than it is for the single electron TB ground state ($\langle r \rangle_{\rm TB} = 1.92$ nm).

The second-electron binding energy BE, corresponding to the energy for the transition $P^{-1} \to P^{0}+ e$, depends on the energy for an unbounded electron in Si or the conduction band edge $E_{\rm CB}$, and is given by

\begin{equation}
  \label{eq:BE1P}
  BE=E_{P^{-1}}-\left(E_{P^0}+ E_{\rm CB} \right).
\end{equation}
Figure \ref{fig:1p2eHF}c shows how the conduction band energy minimum varies with the computational box size. We use these values to estimate the exact value for the $E_{\rm CB}$. Denoting by $x$ the length of the cubic computational box, we anticipate from a particle-in-a-box picture that the conduction band edge will approach as $1/x^2$ to its actual, box-independent value. Using this form, we find that the expression ${\rm CB}(x) = 1.1318$ eV $ + 3.764$eV nm$^2$ $/x^2$ reproduces the trend in Fig.\ \ref{fig:1p2eHF}c, also suggesting $E_{\rm CB} =1.1318$ eV. Utilizing this value for $E_{\rm CB}$, we calculate BE, Eq.\ \eqref{eq:BE1P}, in Fig.\ \ref{fig:1p2eHF}d, and note that the TB-HF calculation predicts a binding energy of -0.93 meV in the largest boxed sized considered. Adding electron correlation energy via the TB-HF-CI$_o$ calculation results in a better estimate of -1.37 meV relative to the experimentally estimated value of -2 meV. We remark that this binding energy is on the order of magnitude of the energies we can compute with the TB-HF method, as numerical convergence is very sensitive to the rapid change in the wavefunction during the self-consistent calculation near the conduction band edge. Nevertheless, this result also shows that one can obtain results closer to the experimental ones for weakly bound states and smaller computational boxes by implementing the HF-TB method.   

\begin{figure}[t]
  \centering
  \includegraphics[scale=0.25]{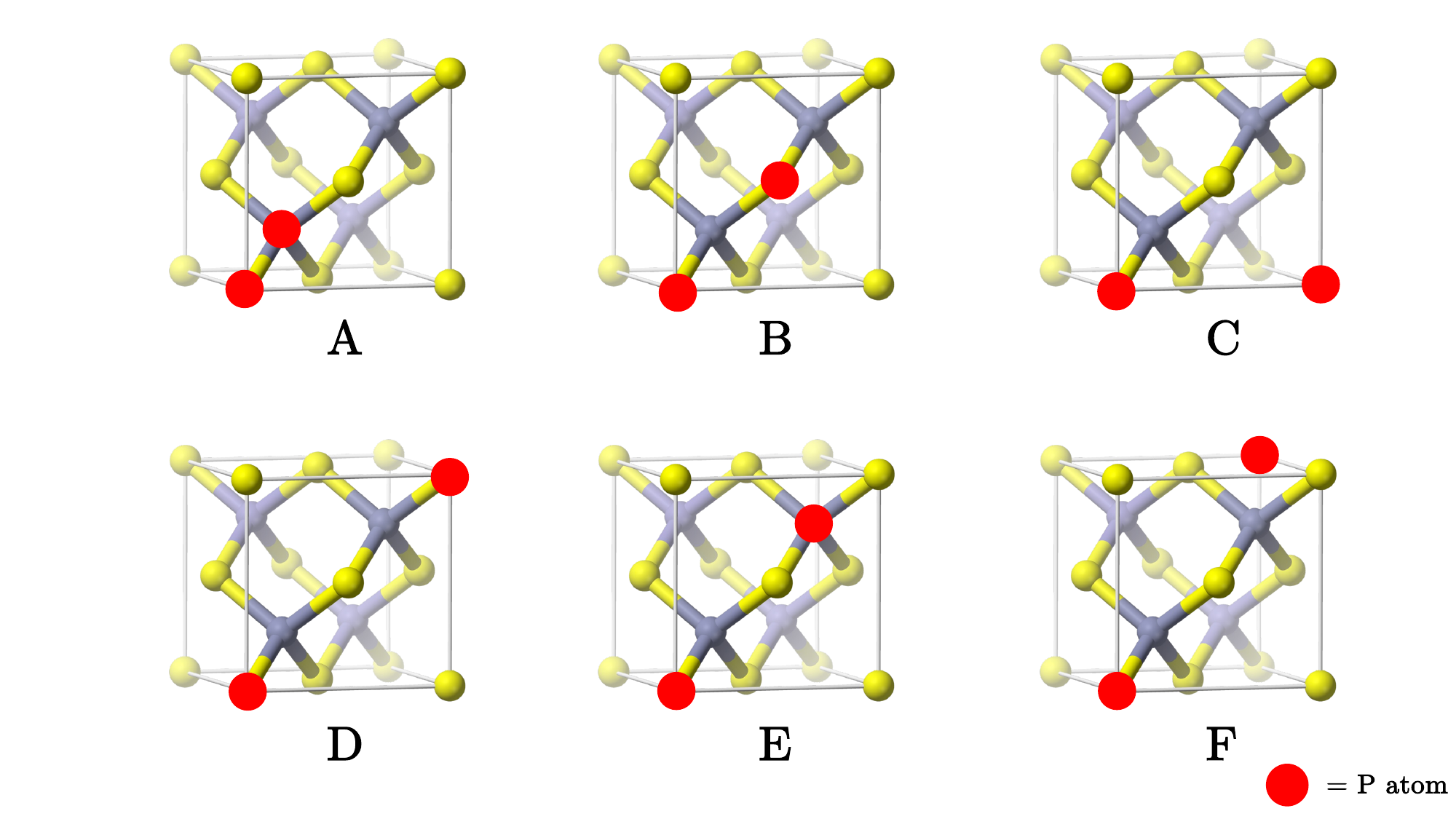}
  \caption{$P_2$ clusters. \textcolor{black}{These systems are embedded in a square computational box of $17.376$ nm edge length in the energy calculations below. }}
  \label{fig:clusters}
\end{figure}

\begin{figure}[t]
  \centering
  \includegraphics[scale=0.85]{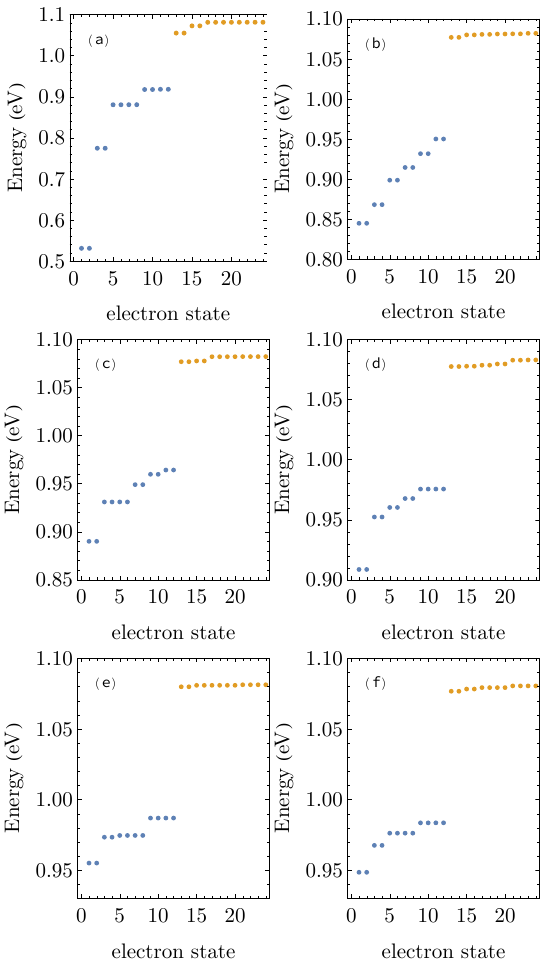}
  \caption{Phosphorus dimer cations P$_2^{+1}$  in silicon. We present single-electron energies for dimers in the configuration one Phosphorous atom in $(0,0,0)$ and the second one at (a) $(1/4,1/4,1/4)$, (b) $(1/2,0,1/2)$, (c) $(1,0,0)$, (d)  $(1,0,1)$, (e) $(3/4,1/4,3/4)$, and (f) $(1,1,1)$. Energies in blue correspond to bound states and those in yellow are conduction band states.}
  \label{fig:2P1eTB}
\end{figure}

\section{Two phosphorus-atom clusters}\label{sec:2Pclusters}

Next, we consider the electronic structure of systems consisting of two close phosphorus atoms in silicon. Specifically, we study the set of structures with two P atoms in Fig.\ \ref{fig:clusters}, which have in common that both phosphorus atoms lay in the same unit cell, forming an atomic cluster. We generically denote these structures by $P_2$ and refer below to each cluster by the label used in Fig.\ \ref{fig:clusters}. These clusters may result from the imperfect fabrication of single-atom transistors or be desired to fine-tune single quantum-dot properties. As shown in Fig.\ \ref{fig:2P1eTB}, the energy spectrum for a single electron in these devices is sensitive to the $P_2$ relative orientation within the silicon matrix and the P-P separation. In contrast with a single $P$ atom, the $P_2$ nuclei form a non-spherical confinement potential. We identify in Fig.\ \ref{fig:2P1eTB} twelve bound states for each case, resulting from valley-splitting and spin degeneracy of the otherwise simple ground-state energy in such non-spherical potential. In contrast with the single-electron energies in a single $P$ atom, the larger electron confinement provided by the $P_2$ cluster shifts the ground state to a lower energy, also increasing the energy range covered by the twelve-state manifold in the Si conduction band gap. Moreover, the non-spherical confinement potential and the relative orientation of the different $P_2$ clusters result in distinct interactions with each conduction band valley. Such differences in the valley splitting manifest in the lower degeneracy in the energy spectrum. As compared with the single $P$ energies\cite{kohn1955theory}, the three-fold $T_2$ states split, forming a two-fold and a one-fold state in clusters A,C,E and F, while all these states show different energies for clusters B and D. Similarly, the two-fold $E$ states are no longer degenerate for the B and C clusters. Thus, single-electron states are very sensitive to the $P_2$ atomic arrangement.

\begin{figure}[t]
  \centering
  \includegraphics[scale=0.77]{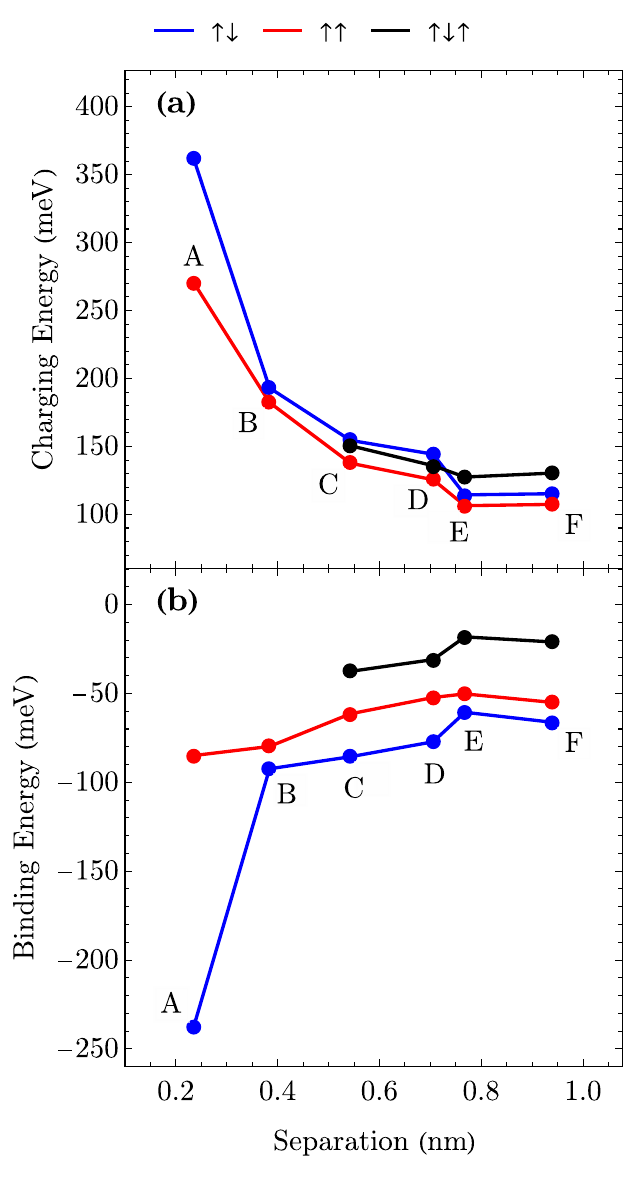}
  \caption{Neutral and anionic phosphorus dimers P$_2^{0}$ and $P_2^{-1}$  in silicon.  (a) Charging energy and (b) binding energy for the second and third electrons. Multielectron configurations: two-electron singlet ($\uparrow \downarrow$, blue), two-electron triplet ($\uparrow \uparrow$, red), and three-electron singlet ($\uparrow \downarrow \uparrow$, black) state. }
  \label{fig:2P2and3ecluster}
\end{figure}

In Fig.\ \ref{fig:2P2and3ecluster} we investigate two and three electron configurations in the $P_2$ clusters in Fig.\ \ref{fig:clusters}. We present the results for the charging and binding energies 
\begin{align}
    BE_n =& E_{P_2^{2-n}} - (E_{P_2^{3-n}} +E_{\rm CB}) \label{eq:BEn}\\
    CE_n =& E_{P_2^{2-n}} - (E_{P_2^{3-n}} +E_{P_2^{+1}}) \label{eq:CEn},
\end{align}
as a function of the impurity separation. For the two-electron configuration $P_2^0$, we analyze the two possible spin-states: a singlet ($\uparrow \downarrow$) and a triplet state ($\uparrow \uparrow$). Comparing the binding energies for each $P_2^0$ spin configuration, we conclude that the singlet state is the ground magnetic state in every case.  We also observe a simple, almost monotonic, trend between the binding energies and the impurity separation in all cases. Charging and binding energies decrease as a function of the $P$-$P$ distance, suggesting that this geometric parameter correlates with the relative electron confinement. Indeed, the A cluster shows the largest second-electron charging and binding energies for both single and triplet states. These energies are significantly smaller, by more than 100 meV in the singlet case, for the B cluster.   However, the variation in the binding energies is less notable between clusters B to F. This suggests that the electronic properties of devices consisting of any of the B-F clusters are similar within a range of tolerance of a few meV if the cluster is charge neutral.  On the other hand, the second electron charging energy is more than twice the corresponding energy in a single impurity. The net result is that these larger charging energies counterbalance the lower single-electron energies observed in the B-F clusters in Fig.\ \ref{fig:2P1eTB}, leading to minor differences in their second-electron binding energy.

\begin{figure}[t]
  \centering
  \includegraphics[scale=1.0]{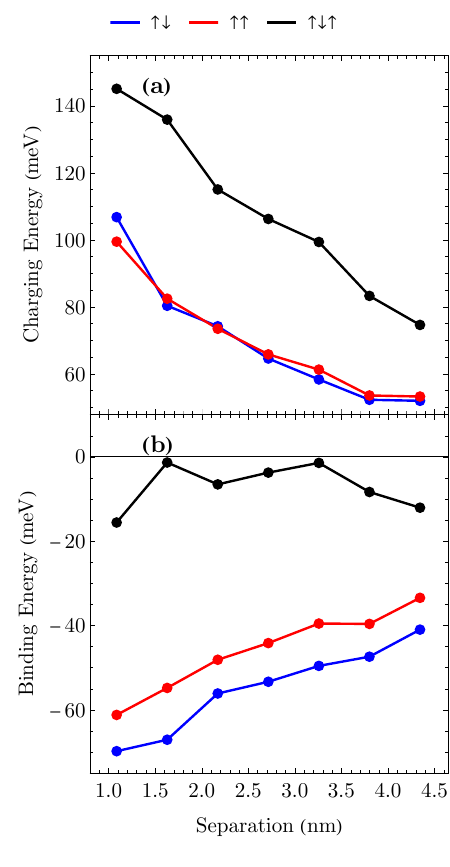}
  \caption{Electron charging and binding energies in a two-phosphorus array as a function of impurity-impurity separation $d$. (a) Charging energy and (b) binding energy for three multielectron configurations: two-electron singlet ($\uparrow \downarrow$, blue), two-electron triplet ($\uparrow \uparrow$, red), and three-electron singlet ($\uparrow \downarrow \uparrow$, black) state. Impurities are separated along the [100] direction.  }
  \label{fig:2Parray}
\end{figure}

\begin{figure}[t]
\centering
  \includegraphics[scale=0.64]{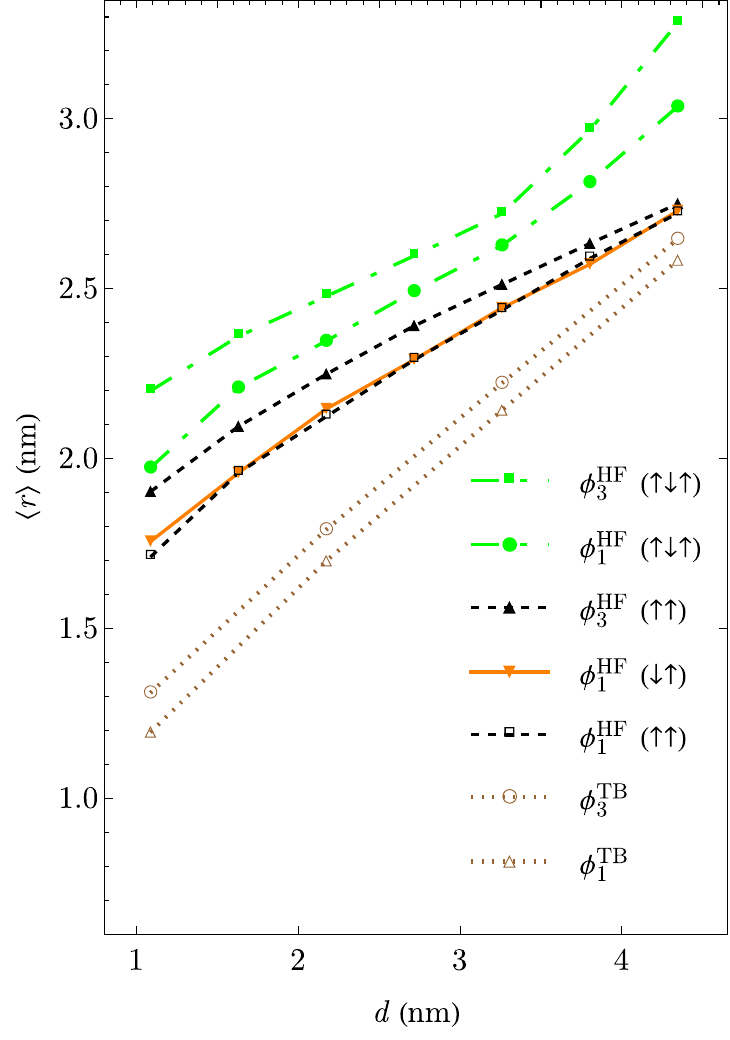}
  \caption{ Electron-density mean radial dispersion  $\langle r \rangle$ in 2P arrays as a function of impurity separation $d$, setting $r_o$ to the midpoint between the impurities.    }
  \label{fig:2PReshape}
\end{figure}

In analogy with the $P^{-1}$ state for a single impurity, we study the $P_2^{-1}$  state for clusters C - F in Fig.\ \ref{fig:clusters}, and report in Fig.\ \ref{fig:2P2and3ecluster} the binding and charging energies for the transition $P_2^{-1} \to P_2^0 + 1 e $. For clusters A and B, our calculations suggest that the $P_2^{-1}$ state is unstable. In our model, these systems are mapped into a three-electron system in the effective potential defined by the $P_2$ cluster. In this form, these systems are analog of the negatively-charged hydrogen molecule $H_2^{-1}$. For the case of $H_2^{-1}$, theoretical studies\cite{norton1976photoconductivity} suggest that the third electron binding does not occur for separations less than 0.15 nm.  For the case of $P_2^{-1}$ in Si, our TB-HF calculation suggests that the third electron binding does not occur below 0.5 nm P-P separation. Moreover, when the $P_2^{-1}$ state is stable, we observe that the charging energy for the third electron is on the order of magnitude of the charging energy for the second electron in clusters C - F. In contrast, the binding energy is smaller than the one for the second electron in $P_2$ but larger than for a second electron in a single P atom.

\section{Two phosphorus-atom arrays}\label{sec:2Parray}

In this section, we study the two and three-electron energies in phosphorus dimers at separations larger than 1 nm. As the distance between P atoms increases, we expect to transition from cluster states $P_2$, corresponding to single quantum dots, to a two-dot array (2P)\cite{ochoa2024single}.

Figure \ref{fig:2Parray} presents the charging and binding energies as a function of P dimer separation between 1 and 4 nm along the [100] direction, calculated from equivalent expressions to those in Eqs.\ \eqref{eq:BEn} and \eqref{eq:CEn}. In this range, for a system with two electrons, the charging energy decays from about 100 meV to nearly 50 meV for both the singlet and triplet spin states. The variation in charging energy is less sensitive to the separation at larger impurity-impurity distance. Binding energies for the second electron also decrease as the P - P distance. Notably, while the charging energy for a second electron is about the same for the singlet and triplet states, the binding energy is larger for the singlet state, implying that the ground state for the neutral dimer 2P is antiferromagnetic. We also note that the binding energy difference between the singlet and triplet states moderately diminishes with separation.

 The binding energy variation as a function of impurity distance for the third electron in the $2P$ arrays in Fig.\ \ref{fig:2Parray}b reveals that this electron is weakly bound, following a nonmonotonic dependence in the range of separations investigated. Indeed, we note in Fig.\ \ref{fig:2Parray}a that the charging energy for a third electron is about 40 \% larger than the corresponding energy for a second electron, driving the binding energy for the third electron in the dimer to lay just below the conduction band edge. The charging energy decays for large P - P separations, favoring a more stable three-electron state. We rationalize this finding in terms of the electron density spread over the confinement regions defined by each phosphorus atom and wavefunction reshaping.  Our calculations confirm that electron density spreads evenly over the two dots, lowering the net electron-electron repulsion at larger separations.  We calculate the mean radial dispersion $\langle r \rangle$, Eq.\ \eqref{eq:mrd}, in Fig.\ \ref{fig:2PReshape}, setting $\vec{r}_o$ to the midpoint in the line separating the two impurities to quantify for wavefunction reshaping. This choice of $\vec{r}_o$ provides a simple interpretation of $\langle r \rangle$ in Fig. \ref{fig:2PReshape} as a local charge density center along the line joining the impurities starting from $\vec{r}_o$. Indeed, if $|\phi(\vec{r})|^2= 1/2 (\delta(\vec{r}-\vec{r}_{P})+\delta(\vec{r}-\vec{r}_{P'})) $, then $\langle r \rangle$  equals half the impurity-impurity distance $ \frac{1}{2}|\vec{r}_P + \vec{r}_{P'}|=d/2$. In all cases analyzed, we observe that $\langle r \rangle > d/2$. Relative to the mean radial distribution for a single electron in the dimer ( states $\phi_1^{\rm TB}$ and $\phi_3^{\rm TB}$ ), we observe in Fig.\ \ref{fig:2PReshape} a significant increase in $\langle r \rangle$ for the HF-states corresponding to the antiferromagnetic ($\downarrow \uparrow$) and ferromagnetic ($\uparrow \uparrow$) two-electron states. The difference in $\langle r \rangle$ between these states narrows for $d > 4$ nm, suggesting that the Coulomb repulsion between the two electrons has a minor effect on the electron distribution beyond this separation.  The HF states for the three-electron configuration ($\uparrow \downarrow \uparrow$) display larger values in $\langle r \rangle$, suggesting that the charge density center near each impurity moves further apart from each other to minimize the electron interactions providing a stable configuration for a third electron.

\section{conclusion}\label{sec:conclusion}
 We introduced the tight-binding Hartree-Fock formalism to calculate many-body properties in dopants in semiconductors. The method relies on atomistic tight-binding calculations of single-electron states, representing the impurities by effective potentials screened by the dielectric constant of the host material. We solve the associated Hartree-Fock equations after obtaining the Coulomb and exchange integrals from numerical solutions to their Poisson equations. The latter can be solved without invoking any particular $sp^3d^5s^*$ atomic orbitals whenever they depend exclusively on the electron density. From the resulting many-body electron energies, we can readily calculate binding and charging energies, which are relevant quantities for understanding and controlling dopant-based devices.  In contrast to the configuration interaction method, the TB-HF protocol presented here provides a more efficient way to account for the reshaping of the electron wavefunction due to confinement and electron interactions. While for many electrons, the computational cost of a CI calculation grows exponentially with the basis size, a typical HF calculation only scales quadratically while depending on several iterations until self-consistency is achieved. Moreover, a CI calculation following a TB-HF calculation allows for additional corrections to the many-body energy due to electron correlation.

\appendix

\begin{acknowledgments}
M.Z. acknowledges support from the Polish National Science Centre based on Decision No. 2015/18/E/STE/00583
 
\end{acknowledgments}

\bibliography{Parray.bib}

\end{document}